\documentclass[structabstract]{aa} % structured abstract

\usepackage{graphicx}
\usepackage{amsmath}
\usepackage{txfonts}
\usepackage{natbib}

\newcommand{\bo}{\ensuremath{\boldsymbol{B}_0}}
\newcommand{\dm}{\ensuremath{D_{\mu\mu}}}
\newcommand{\const}{\ensuremath{\text{const}}}
\newcommand{\abs}[1]{\ensuremath{\left\lvert #1\right\rvert}}

\newcommand{\be}{\begin{equation}}
\newcommand{\ee}{\end{equation}}
\newcommand{\bs}{\begin{subequations}}
\newcommand{\es}{\end{subequations}}

\newcommand{\ez}{\ensuremath{\hat{\boldsymbol{e}}_z}}

\newcommand{\De}{\ensuremath{\varDelta}}
\newcommand{\Om}{\ensuremath{\varOmega}}
\newcommand{\uint}{\ensuremath{\int_{-\infty}^\infty}}
\newcommand{\f}[1]{\ensuremath{\boldsymbol{#1}}}
\newcommand{\m}[1]{\ensuremath{\left\langle #1\right\rangle}}
\newcommand{\pd}[2][]{\ensuremath{\frac{\partial #1}{\partial #2}}}
\newcommand{\dd}[2][]{\ensuremath{\frac{\mathrm{d} #1}{\mathrm{d} #2}}}
\newcommand{\df}{\ensuremath{\mathrm{d}}}

\begin{document}
\title{Pitch-angle scattering in magnetostatic turbulence}
\subtitle{II. Analytical considerations and pitch-angle isotropization}
\author{R.\,C. Tautz}
\institute{Zentrum f\"ur Astronomie und Astrophysik, Technische Universit\"at Berlin, Hardenbergstra\ss{}e 36, D-10623 Berlin, Germany\\\email{rct@gmx.eu}}

\date{Received June 26, 2013; accepted August 21, 2013}

%%%%%
% Abstract: Context (optional), Aims, Methods, Results, Conclusions (optional)
\abstract
{}
{The process of pitch-angle isotropization is important for many applications ranging from diffusive shock acceleration to large-scale cosmic-ray transport. Here, the basic analytical description is revisited on the basis of recent simulation results.}
{Both an analytical and a numerical investigation were undertaken of the Fokker-Planck equation for pitch-angle scattering. Additional test-particle simulations obtained with the help of a Monte-Carlo code were used to verify the conclusions.}
{It is shown that the usual definition of the pitch-angle Fokker-Planck coefficient via the mean-square displacement is flawed. The reason can be traced back to the assumption of homogeneity in time which does not hold for pitch-angle scattering.}
{Calculating the mean free path via the Fokker-Planck coefficient has often proven to give an accurate description. For numerical purposes, accordingly, it is the definition that has to be exchanged in favor of the pitch-angle correlation function.}

\keywords{Plasmas --- Magnetic Fields --- Turbulence --- (Sun:) solar wind --- (ISM:) cosmic rays}
\authorrunning{Tautz et al.}
\titlerunning{Pitch-angle scattering in magnetostatic turbulence}
\maketitle

%%%%%
\section{Introduction}

Evaluating cosmic ray transport in turbulent media, such as the solar wind and the interstellar medium, has been a challenge for nearly half a century \citep[e.g.,][]{par65:pas,jok66:qlt}. Perhaps one of the best known procedures for describing random particle motions is the diffusion-convection class of models \citep{rs:rays,sha09:nli}; later, much attention has been focused on obtaining the forms of the diffusion coefficients from wave-particle interactions. Today, even for spatial diffusion it is not entirely clear that cosmic-ray transport is really diffusive in the sense of a classic Markovian process, for which the diffusion coefficients have finite values in the limit of long times \citep{cha43:sto,tau10:sub}. If a homogeneous background magnetic field is present, however, the dominant process is pitch-angle scattering \citep[see][and references therein]{for77:cor}.

In this series of papers, a systematic investigation has been undertaken of pitch-angle scattering as obtained from analytical predictions of the Fokker-Planck theory and numerical simulations that are based on a Monte-Carlo code \citep{tau10:pad}. In the first paper \citep[hereafter referred to as Paper~I]{tau13:pi1}, pitch-angle scattering of charged energetic particles moving in magnetostatic turbulence was investigated by means of Monte-Carlo test-particle simulations. It was found that, while initially the Fokker-Planck coefficient agrees excellently with quasi-linear \citep[e.g.,][]{jok66:qlt,tau06:sta} and second-order quasi-linear \citep{sha05:soq,tau08:soq} theories, this is not the case for later times. Instead, the Fokker-Planck coefficient---as obtained from the usual prescription $\dm=\langle(\De\mu)^2\rangle/(2t)$ with $\De\mu=\mu(t)-\mu_0$ the pitch-angle displacement---shows a quasi-subdiffusive\footnote{Generally, (spatial) subdiffusion refers to $\langle(\De x)^2\rangle\propto t^\beta$ with $\beta<1$. In contrast, the pitch-angle cosine $\mu$ is bound to the regime $[-1,1]$, which results in a diffusion coefficient that is explicitly time-dependent.} behavior with almost $\dm\propto1/t$. The question had arisen whether such is related to real pitch-angle scattering or, alternatively, whether the basic derivation might be flawed.

Therefore, it is the purpose of this short second paper to shed some light on the theoretical foundations of pitch-angle scattering. Based on the Fokker-Planck equation for pitch-angle scattering, the relation of the pitch-angle mean-square displacement and the Fokker-Planck coefficient for pitch-angle scattering is revisited. The paper is organized as follows. In Sect.~\ref{gen}, the basic relations involving the diffusion coefficient are introduced. In Sects.~\ref{iso} and \ref{sim}, the pitch-angle mean square displacement is derived analytically and compared to numerical simulations, respectively. The modifications for the case of slab scattering are illustrated in Sect.~\ref{slab}. Section~\ref{summ} briefly summarizes the results and discusses the implications.

%%%%%
\section{Pitch-angle diffusion: general considerations}\label{gen}

The Fokker-Planck equation for the ensemble-averaged phase-space distribution function can be derived by transforming the Vlasov equation to the set of gyrocenter coordinates \citep[see][for an introduction]{rs:rays,sha09:nli}. When focusing solely on pitch-angle diffusion, i.e., random variations in the angle $\vartheta=\angle(\f v,\f B_0)$, thus neglecting spatial and momentum diffusion, the differential equation for the particle distribution function \citep{for77:cor} reads as
\be\label{eq:diffMu}
\pd[f(\mu,t)]t=\pd\mu\left(\dm\,\pd[f(\mu,t)]\mu\right)
\ee
with $\mu=\cos\vartheta$. In contrast to the two-dimensional Fokker-Planck equation \citep[e.g.,][]{sha06:2fp,sha09:nli}, spatial homogeneity has been assumed so that $\partial f/\partial z=0$ throughout; here, $z$ is the coordinate along the mean magnetic field $\bo=B_0\ez$. In Eq.~\eqref{eq:diffMu}, the parameter $\dm$ is the Fokker-Planck coefficient, which is defined through the two-time correlation function
\be
\dm(\mu,t)=\int_0^t\df t'\,\m{\dot\mu(t')\dot\mu(0)}
\ee
with the usual, time-independent definition $\dm=\lim\limits_{t\to\infty}\dm(\mu,t)$.

As shown in detail by \citep[Sect.~1.3.2]{sha09:nli}, an equally valid form for the pitch-angle Fokker-Planck coefficient is given through the mean square displacement
\be\label{eq:dmDer}
\dm(\mu,t)=\frac{1}{2}\,\dd t\m{\left(\De\mu\right)^2}
\ee
with $\De\mu(t)=\mu(t)-\mu(0)\equiv\mu-\mu_0$. Central to the mathematical derivation is the assumption of homogeneity in time so that velocity correlation functions only depend on the time difference \citep{sha09:nli}. As becomes clear in the course of the following derivation, this assumption cannot hold for pitch-angle diffusion.

The usual analogy to pitch-angle diffusion is seen in unbound spatial diffusion according to Fick's \citeyearpar{fic55:dif} laws \citep[see also][]{cha43:sto}, where the resulting distribution is
\be
f(x,t)=\frac{1}{\sqrt{4\pi\kappa t}}\,\exp\left[-\frac{\left(x-x_0\right)^2}{4\kappa t}\right].
\ee
By obtaining the second moment,
\be
\m{\left(\De x\right)^2}=\uint\df x\,\left(x-x_0\right)^2f(x,t)=2\kappa t,
\ee
the function $f(x,t)$ gives rise to the famous relation between the diffusion coefficient and the mean square displacement as
\be\label{eq:dx2}
\kappa=\frac{1}{2t}\,\m{\left(\De x\right)^2}.
\ee

A similar definition for the pitch-angle mean-square displacement, i.e., with $\dm$ playing the role of the diffusion coefficient, $\kappa$, seems natural. However, writing
\begin{equation*}
\dm\stackrel{?}{=}\frac{1}{2t}\,\m{\left(\De\mu\right)^2}
\end{equation*}
results in a formula that can, at best, be valid for short times. The same holds true for $\dm$ as defined in Eq.~\eqref{eq:dmDer}, which cannot have a finite, non-vanishing asymptotic value. The reason is that pitch-angle scattering is fundamentally different from spatial diffusion to the limit $\mu\in[-1,1]$. As shown in the next section, the connection between the mean-square displacement and time is not as simple.

%%%%%
\section{Pitch-angle isotropization}\label{iso}

Here, the relaxation process of a distribution of particles  of an initially anisotropic pitch-angle distribution illustrated. The effect of pitch-angle isotropization can be best demonstrated for an ensemble of particles initially with the same pitch-angle cosine, e.g., $\mu_0=0.5$.

\subsection{Distribution function}

Consider again Eq.~\eqref{eq:diffMu}, together with the sharp initial condition that all particles have the same pitch angle, i.e.,
\bs\label{eq:cond}
\be\label{eq:ini}
f(\mu,t=0)=\delta\left(\mu-\mu_0\right),\qquad\mu_0\in(-1,1)
\ee
and the additional requirement that $f$ be normalized, i.e.,
\be\label{eq:constr}
\int_{-1}^1\df\mu\;f(\mu,t)=1\qquad\forall t\geqslant0.
\ee
\es
Equations~\eqref{eq:diffMu} and \eqref{eq:cond} thus represent a partial differential equation with initial condition and integral equality constraint.

There are two simple models for the pitch-angle Fokker-Planck coefficient, which are (i) the isotropic and (ii) the classic slab form. In both cases, the pitch-angle dependence of $\dm$ is grossly simplified. Detailed calculations \citep[e.g.,][]{tau06:sta,sha05:soq,tau08:soq} show that, in fact, the Fokker-Planck coefficient is a complicated form not only of $\mu$ but also of the particle energy and the turbulence power spectrum.

Consider first the isotropic model, where usually
\be\label{eq:DmIso}
\dm=(1-\mu^2)D,\qquad D=\const
\ee
is assumed \citep{sha06:2fp,sha09:nli,ler11:fok}. Then
\be\label{eq:diffMuIso}
\pd[f(\mu,t)]t=D\,\pd\mu\left[\left(1-\mu^2\right)\pd[f(\mu,t)]\mu\right].
\ee
A closed-form analytical solution is not available. However, one can immediately see that the norm is constant by integrating over $\mu$, yielding
\be\label{eq:norm}
\pd t\int_{-1}^1\df\mu\,f(\mu,t)=D\left[\left(1-\mu^2\right)\pd[f(\mu,t)]\mu\right]_{\,\mu=-1}^{\,\mu=1}=0.
\ee
Additionally, a Legendre expansion of the solution has been given by \citet{sha06:2fp} for the sharp initial condition of Eq.~\eqref{eq:ini} as
\be
f(\mu,t)=\frac{1}{2}+\sum_{k=1}^\infty\left(k+\frac{1}{2}\right)P_k(\mu_0)P_k(\mu)\;e^{-k(k+1)Dt}
\ee
with $P_k(z)$ the Legendre polynomial of degree $k$.

\subsection{Pitch-angle Fokker-Planck coefficient}

To determine the mean square displacement, $\langle(\De\mu)^2\rangle$, no knowledge of the solution to the diffusion equation is required. Instead, both the first and the second moments can be directly obtained from the diffusion equation, as is now demonstrated.

%%%%%%%%%%%%%%%%%%%%%%%%%%%%%%%%%%______
\begin{figure}[tb]
\centering
\includegraphics[width=\linewidth]{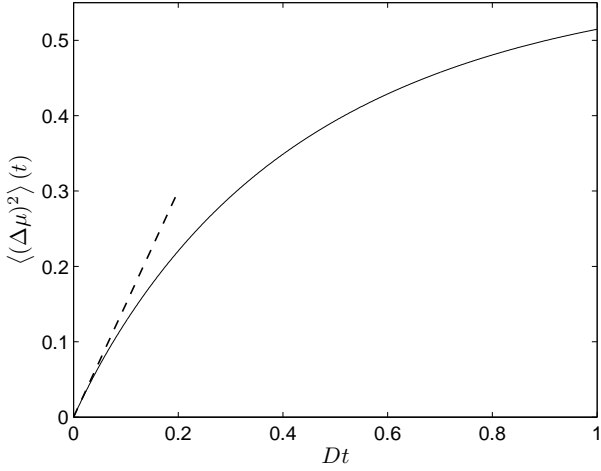}
\caption{Pitch-angle mean-square displacement, $\langle(\De\mu)^2\rangle$, from Eq.~\eqref{eq:msol} for an initial pitch-angle distribution $f(0,\mu)=\delta(\mu-\mu_0)$ with $\mu_0=0.5$. The time is normalized to the constant coefficient, $D$, appearing in Eq.~\eqref{eq:diffMuIso}.}
\label{ab:msqIso}
\end{figure}
%%%%%%%%%%%%%%%%%%%%%%%%%%%%%%%%%%^^^^^^

We commence with the first moment of the distribution function, which enters the derivation of the second moment as shown below. By multiplying Eq.~\eqref{eq:diffMuIso} with $\mu$ and integrating over $\mu$, a differential equation for the first moment is obtained as
\bs
\begin{align}
\pd t\int_{-1}^1\df\mu\;\mu f(\mu,t)&=D\int_{-1}^1\df\mu\;\mu\,\pd\mu\left[\left(1-\mu^2\right)\pd[f(\mu,t)]\mu\right]\\
&=-2D\int_{-1}^1\df\mu\;\mu f(\mu,t),
\end{align}
\es
after repeatedly integrating by parts. The above relation corresponds to
\be
\pd t\m{\mu}=-2D\m{\mu},
\ee
yielding
\be\label{eq:mave}
\m{\mu}(t)=\m{\mu}(t=0)\,e^{-2Dt}.
\ee

Likewise, a differential equation for the second moment $(\De\mu)^2$ can be obtained by multiplying Eq.~\eqref{eq:diffMuIso} with $(\De\mu)^2$ and integrating over $\mu$, yielding
\bs
\begin{align}
&\pd t\int_{-1}^1\df\mu\,\left(\mu-\mu_0\right)^2f(\mu,t)\nonumber\\
=\;&D\int_{-1}^1\df\mu\,\left(\mu-\mu_0\right)^2\pd\mu\left[\left(1-\mu^2\right)\pd[f(\mu,t)]\mu\right]\\
=\;&2D\int_{-1}^1\df\mu\,\left(1-3\mu^2+2\mu\mu_0\right)f(\mu,t), \label{eq:tmp1}
\end{align}
\es
where the righthand side has been integrated by parts (twice). Completing the square, Eq.~\eqref{eq:tmp1} can be rewritten as
\begin{align}
&\pd t\int_{-1}^1\df\mu\,\left(\mu-\mu_0\right)^2f(\mu,t)\nonumber\\
=\;&2D\int_{-1}^1\df\mu\,\left[1+3\mu_0^2-3\left(\mu-\mu_0\right)^2-4\mu\mu_0\right]f(\mu,t).
\end{align}

On the righthand side, the first moment appears, which had been calculated in Eq.~\eqref{eq:mave}. By inserting the normalization condition, Eq.~\eqref{eq:norm}, and by recognizing the definition of the second moment, a differential equation can be obtained as
\be
\pd t\m{\left(\De\mu\right)^2}=2D\left(1+3\mu_0^2-3\m{\left(\De\mu\right)^2}-4\mu_0\m{\mu}\right).
\ee
After inserting the first moment from Eq.~\eqref{eq:mave}, the solution is obtained as
\be\label{eq:msol}
\m{\left(\De\mu\right)^2}=\frac{1}{3}\left(1-e^{-6Dt}\right)+\mu_0^2\left(1+e^{-6Dt}\right)-2\mu_0^2e^{-2dt},
\ee
where, because all particles have pitch-angle $\mu_0$ at time $t=0$, the initial condition is obtained as $\langle(\De\mu)^2\rangle(t=0)=0$. With the time normalized to $D$, the solution is illustrated in Fig.~\ref{ab:msqIso}. It is instructive to note the contrast to Eq.~\eqref{eq:dx2}, which is valid for unbound spatial diffusion.

%%%%%%%%%%%%%%%%%%%%%%%%%%%%%%%%%%______
\begin{figure}[tb]
\centering
\includegraphics[width=\linewidth]{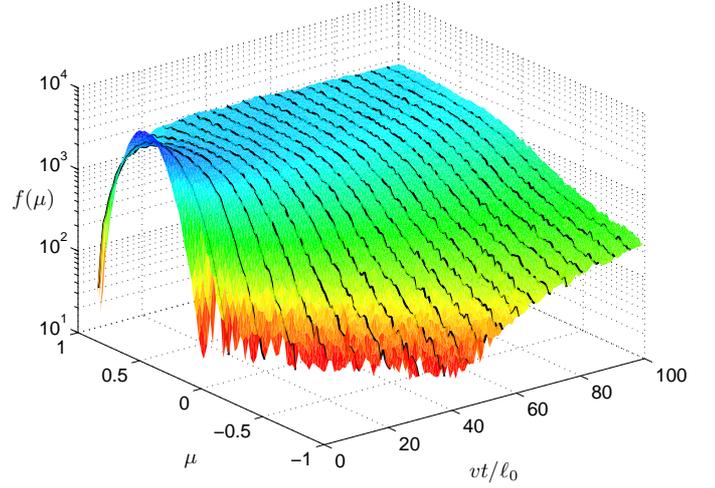}
\caption{(Color online) Isotropization of a particle ensemble in slab turbulence with pitch angles initially fixed at $\mu_0=0.5$. Shown is the time evolution of the pitch-angle distribution.}
\label{ab:muDistW2}
\end{figure}
%%%%%%%%%%%%%%%%%%%%%%%%%%%%%%%%%%^^^^^^

The mean square displacement in Eq.~\eqref{eq:msol} has the following properties:
\begin{itemize}
\item In the limit of long times, one finds that
\be
\lim_{t\to\infty}\m{\left(\De\mu\right)^2}=\frac{1}{3}+\mu_0^2,
\ee
i.e., the mean square displacement is bound by $4/3$;
\item An average mean square displacement can be obtained by integrating over $\mu_0$, yielding
\be
\overline{\m{\left(\De\mu\right)^2}}\equiv\frac{1}{2}\int_{-1}^1\df\mu_0\;\m{\left(\De\mu\right)^2}=\frac{2}{3}\left(1-e^{-2Dt}\right).
\ee
\item For short times, a Taylor expansion of the mean square displacement in Eq.~\eqref{eq:msol} yields
\be
\m{\left(\De\mu\right)^2}=2\left(1-\mu_0^2\right)Dt+\mathcal O(t^2),
\ee
thereby reproducing a quasi-diffusive motion. The factor connecting the mean square displacement and the diffusion coefficient, however, is not merely 2, as was the case for unbound spatial diffusion, but instead involves the initial pitch angle.
\end{itemize}

If one were to rely on Eq.~\eqref{eq:dmDer}, the Fokker-Planck coefficient could then be obtained as
\be
\dm=\frac{1}{2}\,\dd t\m{\left(\De\mu\right)^2}=\left(1-3\mu_0^2\right)De^{-6Dt}+2\mu_0^2De^{-2Dt},
\ee
which obviously tends to zero for long times. For short times, a Taylor expansion yields
\be
\dm\approx\left(1-\mu_0^2\right)D,\qquad t\ll D,
\ee
in accordance with the original assumption. However, as shown in the following section, Eq.~\eqref{eq:dmDer} is \emph{not} the correct formula to connect the pitch-angle mean square displacement and the Fokker-Planck coefficient.

%%%%%
\section{Comparison with numerical simulations}\label{sim}

As seen, the relations $\dm=\tfrac{1}{2}\df\langle(\De\mu)^2\rangle/\df t$ and $\dm=(1-\mu^2)D$, with the requirement that $\dm$ be constant in time, hold only for short time scales. For large time scales, in contrast, $\langle(\De\mu)^2\rangle$ should become asymptotically constant so that, according to the usual definition, the Fokker-Planck coefficient tends to zero with a time dependence $1/t$. As shown in Paper~I, however, one observes that particles still undergo pitch-angle scattering even for long times.

A solution to this dilemma may be found in a purely simulational approach, which involves the distribution function itself as obtained from a Monte-Carlo simulation. Such simulations \citep[see, e.g.,][]{tau10:pad,tau13:num} trace the trajectories of a large number of test particles moving in a given turbulent magnetic field structure. A more detailed description of the code can be found in Paper~I, where extensive use has been made of such simulations. It is a simple matter to obtain the time evolution of the pitch-angle distribution function as shown in Fig.~\ref{ab:muDistW2} for slab turbulence. As shown, an initial distribution of $f(\mu,0)=\delta(\mu-\mu_0)$ will quickly become isotropized due to extensive pitch-angle scattering (also and especially through $90^\circ$). The time is normalized to the particle velocity $v$ and the turbulence bend-over scale $\ell_0$ so that $vt/\ell_0$ is dimensionless.

It therefore should be possible to obtain the Fokker-Planck coefficient directly by integrating the diffusion equation~\eqref{eq:diffMu} so that
\be
\dm(\mu,t)=\left(\pd[f(\mu,t)]\mu\right)^{-1}\pd t\int_{-1}^\mu\df\mu'\;f(\mu',t).
\ee
In practice, however, it turns out that, even for a simulation with as many as $10^7$~particles, extensive filtering and/or smoothing would be required, which due to the derivative in the denominator, renders the results invalid.

%%%%%%%%%%%%%%%%%%%%%%%%%%%%%%%%%%______
\begin{figure}[tb]
\centering
\includegraphics[width=\linewidth]{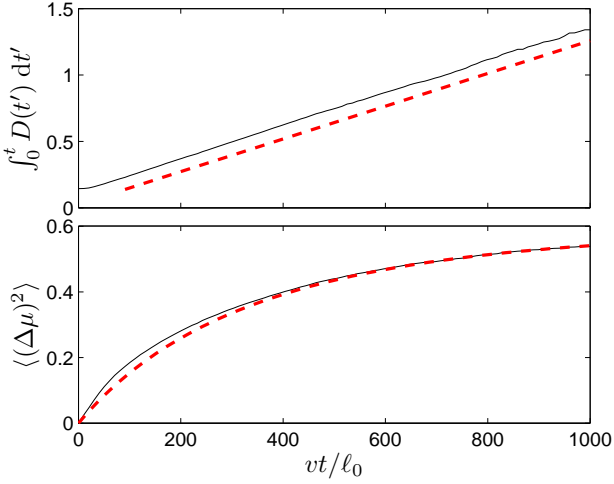}
\caption{(Color online) Upper panel: Integral of the isotropic scattering coefficient, $D$, as obtained from a numerical simulation (black solid line) for particles with a normalized rigidity $R=\gamma v/(\Om\ell_0)=0.1$. A linear fit (red dashed line) yields the value for $D=1.2324\times10^{-3}\Om$ itself; for clarity, the fit line has been shifted downward. Lower panel: Pitch-angle mean square displacement as obtained from the simulation (black solid line) in comparison with the analytical form (red dashed line) from Eq.~\eqref{eq:msol}, which uses the value for $D$ that was obtained above.}
\label{ab:CorMsq}
\end{figure}
%%%%%%%%%%%%%%%%%%%%%%%%%%%%%%%%%%^^^^^^

Therefore, another approach is required. Because of $v_z(t)=v\mu(t)$, the parallel velocity auto-correlation function can be expressed according to \citet{sha12:nd1} as
\be
\m{v_z(t)v_z(0)}=\frac{v^2}{3}\,\exp\left[-2\int_0^t\df t'\;D(t')\right]
\ee
with $D$ the (now time-dependent) coefficient in Eq.~\eqref{eq:diffMuIso}. We write
\be\label{eq:Dcorr}
\int_0^t\df t'\;D(t')=-\frac{1}{2}\ln\left(\frac{3\m{v_z(t)v_z(0)}}{v^2}\right)
\ee
so that $D(t)$ corresponds to the slope of the righthand side. A test-particle simulation confirms that, for isotropic scattering and particles starting initially with a fixed pitch angle, $D$ is indeed a constant (see upper panel in Fig.~\ref{ab:CorMsq}). This is true at least for times that are significantly longer than for the pitch-angle mean-square displacement; eventually $D$ will vanish owing to $\mu(t)\propto v_z(t)$. As has been shown previously (and is clear from intuition), the pitch-angle auto-correlation function will turn to zero, too \citep{fra12:aut}.

As a second step, the resulting $D$ can be plugged into the analytical solution for the pitch-angle mean-square displacement, Eq.~\eqref{eq:msol}. This allows for a comparison with $\langle(\De\mu)^2\rangle$ as obtained directly from a numerical test-particle simulation. This confirms the important result of Sect.~\ref{iso} that the pitch-angle mean-square displacement does not grow linearly with time as required for a diffusion approach.

%%%%%
\section{Classic slab scattering}\label{slab}

%%%%%%%%%%%%%%%%%%%%%%%%%%%%%%%%%%______
\begin{figure}[tb]
\centering
\includegraphics[bb=110 195 480 645,clip,width=\linewidth]{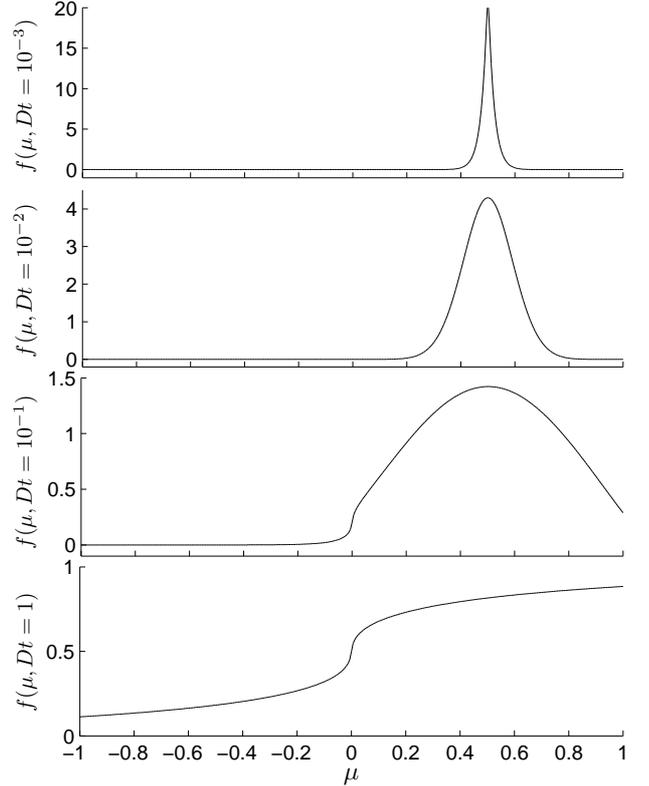}
\caption{Solution to the pitch-angle diffusion equation, Eq.~\eqref{eq:diffMu} for slab scattering with $s=5/3$. The initial condition has been chosen according to Eq.~\eqref{eq:ini} with $\mu_0=0.5$.}
\label{ab:distriSlab}
\end{figure}
%%%%%%%%%%%%%%%%%%%%%%%%%%%%%%%%%%^^^^^^

In contrast to the isotropic model, where $\dm=(1-\mu^2)D$, the slab model assumes
\be
\dm=(1-\mu^2)\abs\mu^{s-1}D,
\ee
where $s=5/3$ is the (Kolmogorov) index of the magnetic turbulence power spectrum \citep[e.g.,][]{tau06:sta,sha06:2fp}.

In Fig.~\ref{ab:distriSlab}, the evolution of the pitch-angle distribution is shown at four different times. Especially the third panel underscores that, due to the additional factor $\lvert\mu\rvert^{s-1}$, which is responsible for $\dm(\mu=0)=0$, pitch-angle scattering through $90^\circ$ is severely impeded. However, even though analytically impossible, the last panel shows that a certain fraction of the particles is eventually scattered backward, which can be understood in terms of numerical diffusion.

\begin{figure}[tb]
\centering
\includegraphics[width=\linewidth]{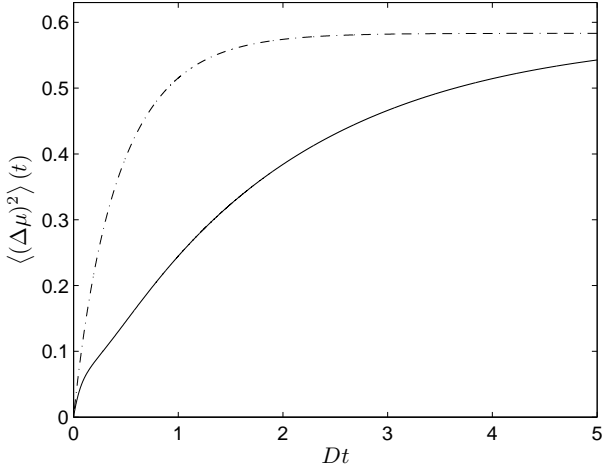}
\caption{Pitch-angle mean square displacement (solid line) for slab scattering as obtained from the numerical solution of Eq.~\eqref{eq:diffMu} with $\dm=(1-\mu^2)\lvert\mu\rvert^{s-1}D$. The result for isotropic scattering from Eq.~\eqref{eq:msol} is shown for comparison (dot-dashed line; see Fig~\ref{ab:msqIso}).}
\label{ab:msqSlab}
\end{figure}

The ineffectiveness of $90^\circ$ pitch-angle scattering is also reflected in the mean square displacement, which, when compared to the result for isotropic scattering, is significantly lower as shown in Fig.~\ref{ab:msqSlab}. Owing to the fractional $\mu$ dependence of the Fokker-Planck coefficient, however, it is no longer feasible to obtain an analytical expression solely from the diffusion equation.

%%%%%
\section{Discussion and conclusion}\label{summ}

In this series of papers, random variations in the pitch-angle of charged particles have been investigated that move in a turbulent magnetic field. In Paper~I, the case of an initially isotropic pitch-angle distribution was investigated. There, the question is pursued as to whether and to what extent the numerically obtained Fokker-Planck coefficient agrees with analytical descriptions. It was found that, for long times, the numerically obtained Fokker-Planck coefficient of pitch-angle scattering tends to zero with a $1/t$ behavior.

%%%%%%%%%%%%%%%%%%%%%%%%%%%%%%%%%%______
%%%%%%%%%%%%%%%%%%%%%%%%%%%%%%%%%%^^^^^^

It was the purpose of this second paper to trace the aforementioned behavior back to the Fokker-Planck equation, where the coefficient had originally been introduced in the same style as the spatial diffusion coefficient. However, as shown in Sect.~\ref{iso}, the mean square displacements for pitch-angle scattering and for spatial diffusion behave entirely differently due to the limitation $\mu\in[-1,1]$. Accordingly, the homogeneity in time is no longer fulfilled, an assumption however that is central for the derivation \citep[see][Sect.~1.3.2]{sha09:nli} of Eq.~\eqref{eq:dmDer}.

Instead, the mean-square displacement as obtained from the parallel velocity auto-correlation function has proven to be in almost perfect agreement with the results from a test-particle simulation. Therefore, the parallel velocity auto-correlation function should be more suitable to obtain a Fokker-Planck coefficient. Combine Eqs.~\eqref{eq:Dcorr} and \eqref{eq:DmIso} to obtain
\be\label{eq:dmCor}
\dm=-\frac{1}{2}\left(1-\mu^2\right)\dd t\,\ln\left(\m{\mu(t)\mu(0)}\right),
\ee
thus allowing use of the slope of the resulting curve.

If the correlation function decays exponentially, i.e., $\langle\mu(t)\mu(0)\rangle\propto e^{-Dt}$ with $D=\const$, then $\dm$ will attain a finite value within the limit of long times. Both Fig.~\ref{ab:CorMsq} and earlier results \citep[e.g.,][even though it is argued that the decay proceeds significantly slower than expected]{fra12:aut} confirm an exponentially decaying pitch-angle correlation function. Numerically, Eq.~\eqref{eq:dmCor} will eventually become zero due to numerical round-off errors. However, as illustrated in Fig.~\ref{ab:CorMsq}, the range of validity for the parallel velocity auto-correlation function, i.e., the time period for which $D\approx\const$, seems to be wider than for the pitch-angle mean-square displacement.

Summarizing, the Fokker-Planck coefficient $\dm$ is required for multiple applications, including the evaluation of spacecraft data and the theoretical determination of the mean-free path. In Paper~I it has been shown that, in contrast to theoretical expectations, $\dm$ tends to zero for late times. Here it has been explained by demonstrating that, to calculate \dm, the pitch-angle correlation function is to be preferred over the pitch-angle mean-square displacement. Furthermore, the mean free path can be obtained directly from the parallel velocity auto-correlation function without invoking the pitch-angle Fokker-Planck coefficient \citep[see][]{sha12:nd1}. As shown in Paper~I, the Fokker-Planck coefficient is non-zero even for a relaxed pitch-angle distribution since the particles are still being scattered. Therefore, future work needs to revisit the pitch-angle scattering of real particles as obtained from spacecraft data, which in most cases relies on the (questionable) use of the pitch-angle mean-square displacement.

%%%%%
\begin{acknowledgements}
I acknowledge useful discussions with Horst Fichtner, Andreas Kopp, Frederic Effenberger, Alexander Dosch, and Ian Lerche. Furthermore, I thank Jan Bolte for help with the numerical scheme I used for solving the partial differential equations with integral constraints.
\end{acknowledgements}

%%%%% Bibliography BEGIN
% \bibliography{../../tautz,../../book,../../article}
% \bibliographystyle{aa}

%%%%% Bibliography END

\end{document}